\documentclass[singlecolumn]{aastex7}
\usepackage{graphicx}

\usepackage{color}

\begin{document}

\title{Investigation of magnetic topology and triggering mechanisms of a C-class flare and active-region blowout jet.}

\author[orcid=0000-0001-8200-4053]{Yogesh Kumar Maurya}
\affiliation{Udaipur Solar Observatory/ Physical Research laboratory, Ahmedabad, India.}
\email[show]{yogeshjn1995@gmail.com}  

\author[orcid=0000-0003-4522-5070]{Ramit Bhattacharyya} 
\affiliation{Udaipur Solar Observatory/ Physical Research laboratory, Ahmedabad, India.}
\email{ramit73@gmail.com}

\author[orcid=0000-0002-6442-7818]{Peter Wyper} 
\affiliation{Department of Mathematical Sciences, Durham University, Durham, DH1 3LE, UK}
\email{peter.f.wyper@durham.ac.uk}

\begin{abstract}
Coronal jets are collimated plasma eruptions which are ubiquitous in the solar atmosphere. Believed to be triggered by magnetic reconnection, these jets can contribute to various phenomena, including coronal heating and particle acceleration. Coronal jets are a contemporary area of research with their onset mechanism meriting further attention. Importantly, a subclass of jets, the blowout jets, are particularly interesting because of their broad spire, suggesting substantial three-dimensional (3D) reconnection between open and closed field lines involving 3D null points. Consequently, here we explore the onset of a blowout jet associated with Active Region (AR) SPoCA 29093 detected by Spatial Possibilistic Clustering Algorithm (SPoCA). This AR produced a C$1.1$-class flare on $10$ November $2022$ and we investigate it using a data-constrained magnetohydrodynamic simulation initiated with a non force-free-field (NFFF) extrapolation of the photospheric magnetic field. Key elements of the extrapolated field lines are the presence of a 3D null and a magnetic flux rope (MFR) co-located with the jet activity region, the evolution of which is further traced in the simulation. The simulation suggests that magnetic reconnection is responsible for the evolution of the MFR, leading to a near-simultaneous onset of the flare and jet as observed by the AIA/SDO. In particular, the simulation shows spontaneous creation and annihilation of 3D null pairs via magnetic reconnection near the jet region. Such spontaneous null pair generation, in principle, can trigger or contribute to coronal jets; opening up a new avenue for further research. 

\end{abstract}



\section{Introduction}
Solar jets are narrow, collimated, transient plasma ejections observed throughout the solar atmosphere, occurring in diverse magnetic environments ranging from active regions to coronal holes \citep{2019ApJ...885L..15K, 2013ApJ...769L..21A, 2018ApJ...854..155K, 2019ApJ...873...93K, 2016ApJ...832L...7P}. They span a wide range of spatial and temporal scales \citep{2007Sci...318.1591S, 2016SSRv..201....1R} and are observed in multiple phenomena like chromospheric surges and spicules \citep{2003A&A...402..361T, 2007Sci...318.1574D, 2007Sci...318.1580C} to coronal X-ray and EUV jets \citep{1992PASJ...44L.173S, 1998ASPC..154..329G, 2009A&A...508.1443B}. Common to all jets are their impulsive onset, association with magnetic structures, and rapid outward flows. Occasionally, they can have additional helical or rotational motions \citep{2004ApJ...610.1129J, 2008ApJ...680L..73P, 2009SoPh..259...87N} but always show transient brightenings, suggesting magnetic reconnection to be their underlying cause \citep{2008ApJ...673L.211M, 2004GApFD..98..407P, 2009PhPl...16l2101P}. This transient brightenings can further manifest as microflares and bright points \citep{1996PASJ...48..123S, 2000ApJ...542.1100S}.

Coronal jets can be classified into two main categories: a standard jet and a blowout jet. Standard jets typically have an inverted Y-shape. Classically, these jets were thought to be produced by reconnection between closed and open field lines occurring at a current sheet surrounding at least one 3D null point. Reconnected open field lines then guide plasma along a spire which is significantly narrower than its base \citep[e.g.][]{1992PASJ...44L.173S}. 
 
Contrarily, blowout jets have wide spires, comparable in width to their bright bases \citep{2010ApJ...720..757M, 2013ApJ...769..134M}. These jets were initially thought to be initiated by a different reconnection scenario, i.e., the eruption of a small-scale flux rope or mini-filament that blows out the overlying magnetic canopy near a 3D null point producing broader spires having more complex magnetic structure and can lead to larger mass ejections \citep{2010ApJ...720..757M, 2013ApJ...769..134M}. However, more recent work has suggested that some (or even all) standard jets may also arise from such eruptions \citep[e.g.][]{2015Natur.523..437S}.

Many of the physical and observational characteristics of coronal jets have been studied with Magnetohydrodynamic (MHD) simulations \citep[e.g.][]{1995Natur.375...42Y, 1996PASJ...48..353Y, 2004ApJ...614.1042M, 2005ApJ...635.1299A, 2007A&A...466..367A, 2010A&A...512L...2A} including the generation of thermal plasma flow---producing jet emissions through reconnection driven heating \citep{2009ApJ...691...61P, 2009ApJ...704..485T}. For instance, \citet{2013A&A...555A.123B} have shown that reconnection-driven heating in their model reproduced the observed 3D structure of coronal loops with plasma flows along them, demonstrating that MHD models are accurate enough to infer the coronal magnetic field configuration on observable scales.

A widely accepted theoretical and observational framework for both types of jet based on a series of simulations and observational works is the magnetic breakout reconnection model \citep{2017Natur.544..452W, 2018ApJ...852...98W, 2019ApJ...887..246D}. In this model, an embedded bipole in a fan–spine magnetic topology reconnects with surrounding fields, removing the overlying magnetic field lines and allowing the low-lying core field to rise and, subsequently, erupt. The model unifies the physics of coronal jets, confined eruptions, and large-scale coronal mass ejections within a common reconnection-driven scenario \citep{2018ApJ...852...98W, 2017Natur.544..452W, 2015Natur.523..437S}. Possible ways to initiate reconnection in the breakout model can be through instability, flux emergence, or shear flows at the polarity inversion line \citep[e.g.][]{Wyper2019, 2019ApJ...885L..15K, 2023ApJ...943..156K}.

Despite the attention coronal jets have received, the magnetic field evolution and trigger mechanism of blowout jets in particular are often challenging to reproduce and interpret in given events because of the involved complexity of magnetic field lines (MFLs), and merit further attention. In this work we conduct a detailed exploration of the reconnection and magnetic topology evolution in a blowout jet using a data-constrained simulation combined with multi-wavelength observations. The observed event was a C$1.1$-class flare and blowout jet associated with Active Region (AR) SPoCA 29093 detected by Spatial Possibilistic Clustering Algorithm (SPoCA) \citep{2014A&A...561A..29V} on November $10, 2022$ at $03:12:00$ UT and is selected since, to the best of our knowledge, the event has not been studied before. The multi-wavelength observations are from the Atmospheric Imaging Assembly (AIA) onboard the Solar Dynamics Observatory (SDO). A coronal field extrapolation is carried out using a NFFF extrapolation from a vector magnetogram from HMI/SDO, while for the simulation, a 3D model, EULAG-MHD, is employed.

The paper is organized as follows: Section~\ref{sec:event_obs} describes the event observations; Section~\ref{sec:NFFF} presents the non-force free field extrapolation model and the important properties associated with a 3D null point; Section~\ref{sec:numerical_model_setup} outlines the MHD simulation model and setup; Section~\ref{sec:MHD_simulations} reports the results; and Section~\ref{sec:summary} summarizes and discusses our findings.

\section{Event observation and its details} \label{sec:event_obs}

\begin{figure}[ht]
\includegraphics[width=\textwidth]{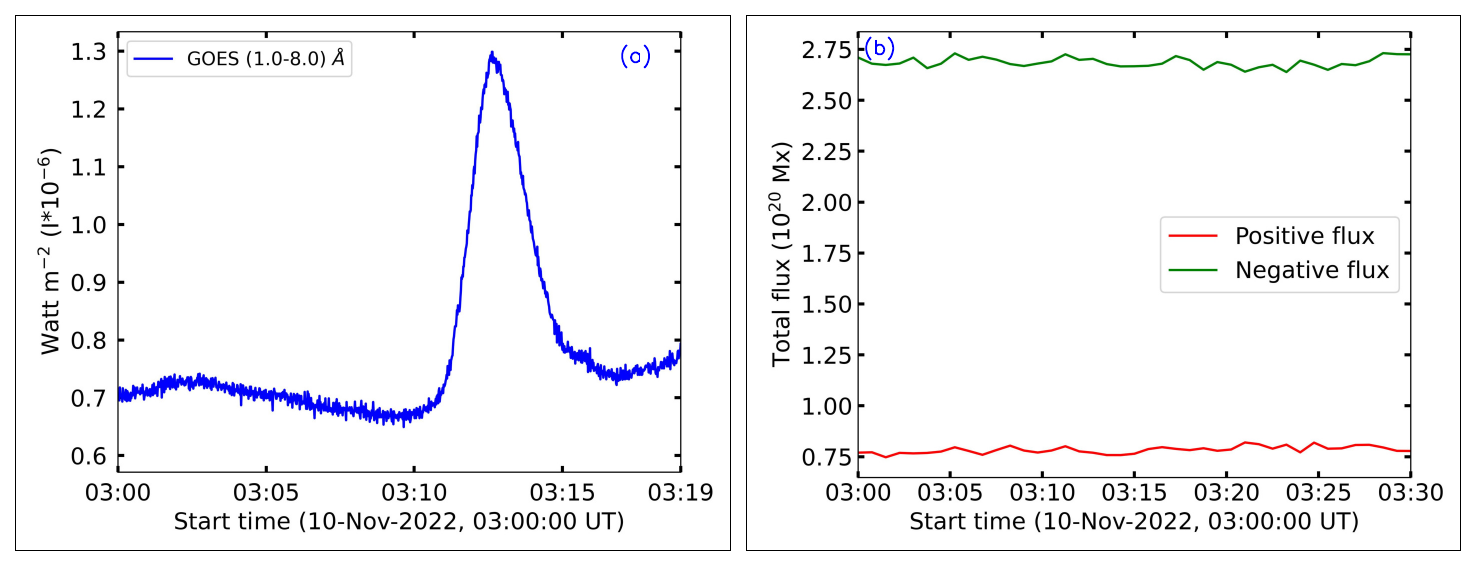}
\caption{The panel (a) depicts GOES ($1-8$) \AA~ channel soft X-ray flux variation (blue curve) for about $19$ minutes starting from $03:00:00$ UT on November $10, 2022$. The flare starts at $03:09:00$ UT followed by a decrease after a peak at $03:12:00$ UT, resulting in a C $1.1$ class flare. The total positive and negative line-of-sight photospheric magnetic flux (red and green solid curves, respectively) remains constant for about 30 minutes, suggesting no significant flux change during the flare and jet (panel (b)). The change in positive and negative flux is well within $1\%$ and $0.5\%$, respectively.}
\label{fig:goes_plot_and_los_flux}
\end{figure}

\begin{figure}
\includegraphics[width=\textwidth]{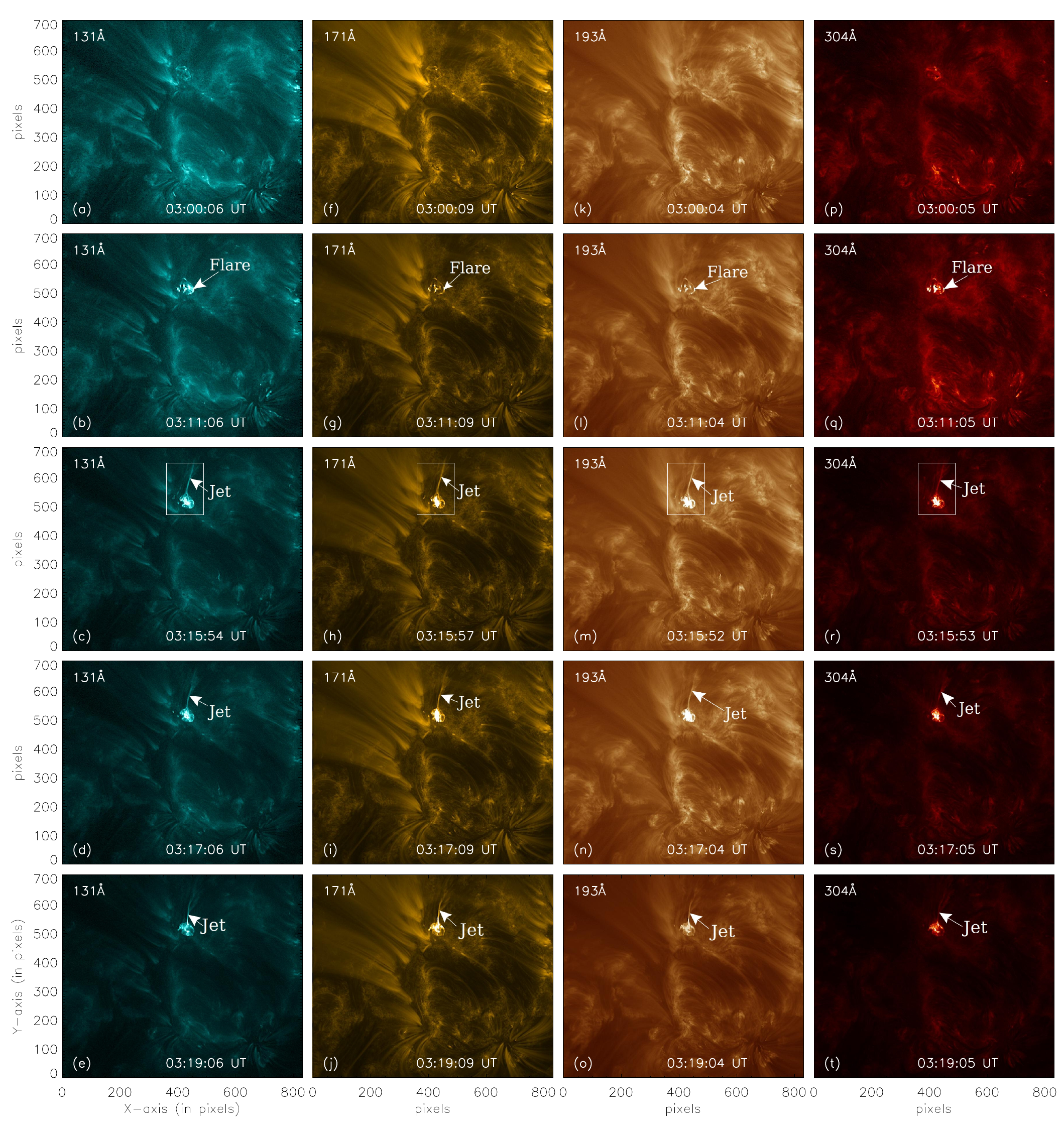}
\caption{Figure depicts the evolution of a C $1.1$ class circular ribbon flare and a blowout jet activity using four AIA intensity channels (column-wise). Each panel spans over ($301.6 \times 255.2$) Mm covered with ($832 \times 704$) pixels in the x and y-axes, respectively. The evolution of flare and jet activity is shown in the first column (from left) for AIA 131 \AA, in the second column for AIA 171 \AA, in the third column for AIA 193 \AA, and in the last column for AIA 304 \AA. The first row (from top) depicts the observation at $03:00$ UT with no prominent activity, the second row represents the C $1.1$ class flare at $03:11$ UT and a jet around $03:16$ UT (shown in the third row) with the white box representing the ROI. The fourth and fifth rows represent the evolution of the jet around $03:17$ UT and $03:18$ UT, respectively. The Corresponding animation is available (in $2 \times 2$ panels) in the final HTML version with $t\in\{03:00, 03:19\}$ UT, where the flares and blowout jet can be better visualized in four AIA channels. The real time of the animation is $12$ s}.
\label{fig:all_aia_channels}
\end{figure}

This event was selected because of (a) its proximity to the solar disk center, which ensures minimal errors in the observed photospheric magnetic field; (b) the photospheric magnetic flux integrated across the active region remains approximately constant for half an hour covering the flare and jet activity, allowing one to explore the mechanism other than flux emergence responsible for blowout jet; (c) the multiwavelength observational nature of the event. Figure \ref{fig:goes_plot_and_los_flux}(a) shows the variation in the Geostationary Operational Environmental Satellite (GOES) soft X-ray flux in the $1-8$ \AA~ channel for the duration of $19$ minutes starting from $03:00$ UT, covering the flare and onset of the jet. The vertical axis represents GOES flux (I) scaled by $10^{-6}$ in the unit of watt m$^{-2}$. The flare starts at $03:09:00$ UT and the peak is around $03:12:00$ UT, followed by a decrease. Panel (b) depicts the evolution of total positive (red curve) and negative (green curve) line-of-sight (LOS) photospheric magnetic flux obtained from the $hmi.M45$ data series of HMI \citep{2012SoPh..275..229S, 2012SoPh..275..207S} for region of interest (ROI) encompassing the flare and jet activity region and found to be nearly constant, varying only $1$ \% and $0.5$\% WRT their initial values. The ROI spans over ($47.125 \times 65.250$) Mm covered with ($130 \times 180$) pixels along x and y, respectively and is highlighted by the white rectangular box in panel (c) of Fig. \ref{fig:all_aia_channels}. 

The multiwavelength evolution of the flare and jet event using AIA observations is shown in Figure \ref{fig:all_aia_channels}. The evolution covers around $19$ minutes starting from $03:00$ UT. An animation (movie-1a, 1b, 1c, 1d) is provided for the Fig. \ref{fig:all_aia_channels} which shows the continuous evolution of flare and jet activity and spans $19$ minutes across AIA channels. The size of each panel is ($301.6 \times 255.2$) Mm covered with ($832 \times 704 $) pixels along the X- and Y- directions, respectively. The event has been observed in all AIA channels, but the shown AIA channels (131 \AA , 171 \AA , 193 \AA , and 304 \AA ) are selected for the demonstration of the event and to show its multi-wavelength nature. The first row depicts observation around $03:00$ UT in four AIA channels with no prominent activity seen (panels (a), (f), (k) and (p)) and last row represents the observation around $03:18$ UT when a blowout jet has been fully developed (panels (e), (j), (o) and (t)). The second row represents the observation around $03:11$ which marks a C $1.1$ class flare (panels (b), (g), (l), and (q)) and third row depicts the observation of jet (highlighted by a rectangular white box) around $03:16$ UT (panels (c), (g), (m), and (r)). Further evolution of the jet around $03:17$ UT is shown in the fourth and fifth rows.

\begin{figure}
\includegraphics[width=0.859\textwidth]{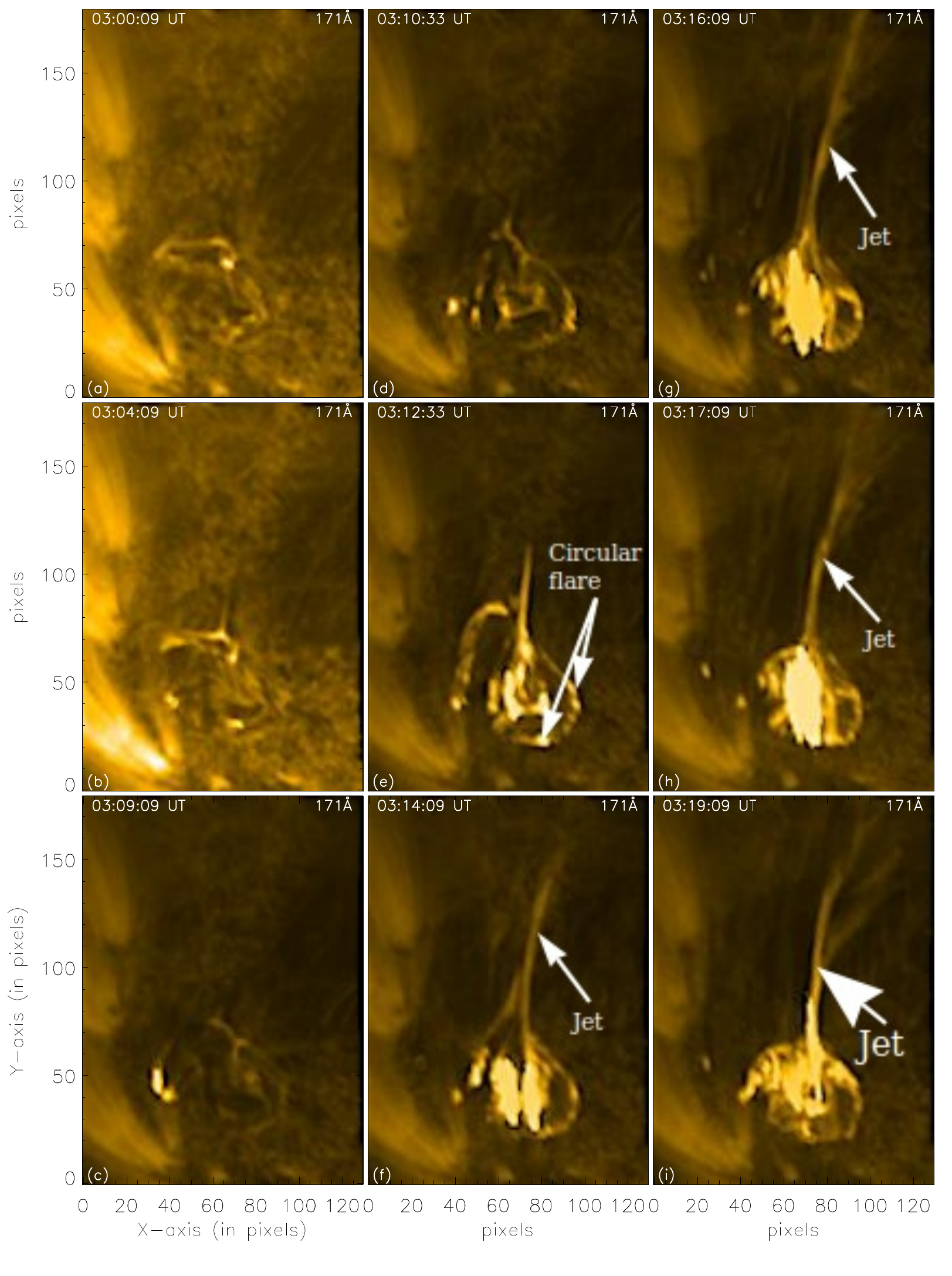}
\caption{Figure depicts the detailed evolution of flare and jet using AIA $171$ \AA~channel, considering the same ROI. Initially, no significant activity is observed (panel (a)), and subsequently, pre-flare brightening starts to occur (panels (b)-(d)), followed by a C $1.1$ class flare prominent in panel (e) and a blowout jet (panel (f)-(i)). The corresponding animation is available in the final HTML version with $t\in\{03:00, 03:19\}$ UT, where the pre-flare brightening, flares, and a blowout jet can be better visualized. The real time of the animation is $18$ s.}
\label{fig:single_evolution}
\end{figure}

To further explore the evolution of flare and jet in detail, the same ROI is plotted in Fig. \ref{fig:single_evolution} using AIA $171$ \AA~ observation based on its better visuals of the event. Panel (a) depicts no significant brightening present at $03:00$ UT. The initial brightening before the flare is shown in panels (b)-(d). Notably, subsequent evolution shows a C $1.1$ class circular flare around $03:12:33$ UT (panel (e)) and a blowout jet onset around $03:14$ UT (panel (f)). The jet evolution is shown by panels (g)-(i), covering time $\in\{03:16$ UT to $03:19\}$ UT. The continuous evolution can be seen in the provided accompanying animation (movie-2). To explore magnetic field topology, dynamics associated with flare, and the onset mechanism of blowout jet, a data-constrained magnetohydrodynamics (MHD) simulation is performed for the same time span as the animation: $19$ minutes starting from $03:00$ UT. The initial coronal magnetic field at $03:00$ UT is constructed by employing the non-force-free field extrapolation technique. The details of the NFFF extrapolation model are discussed in the next section (Sect. \ref{sec:NFFF}).

\section{Non-Force Free Field (NFFF) Extrapolation model}\label{sec:NFFF}

\begin{figure}
\includegraphics[width=1.0\textwidth]{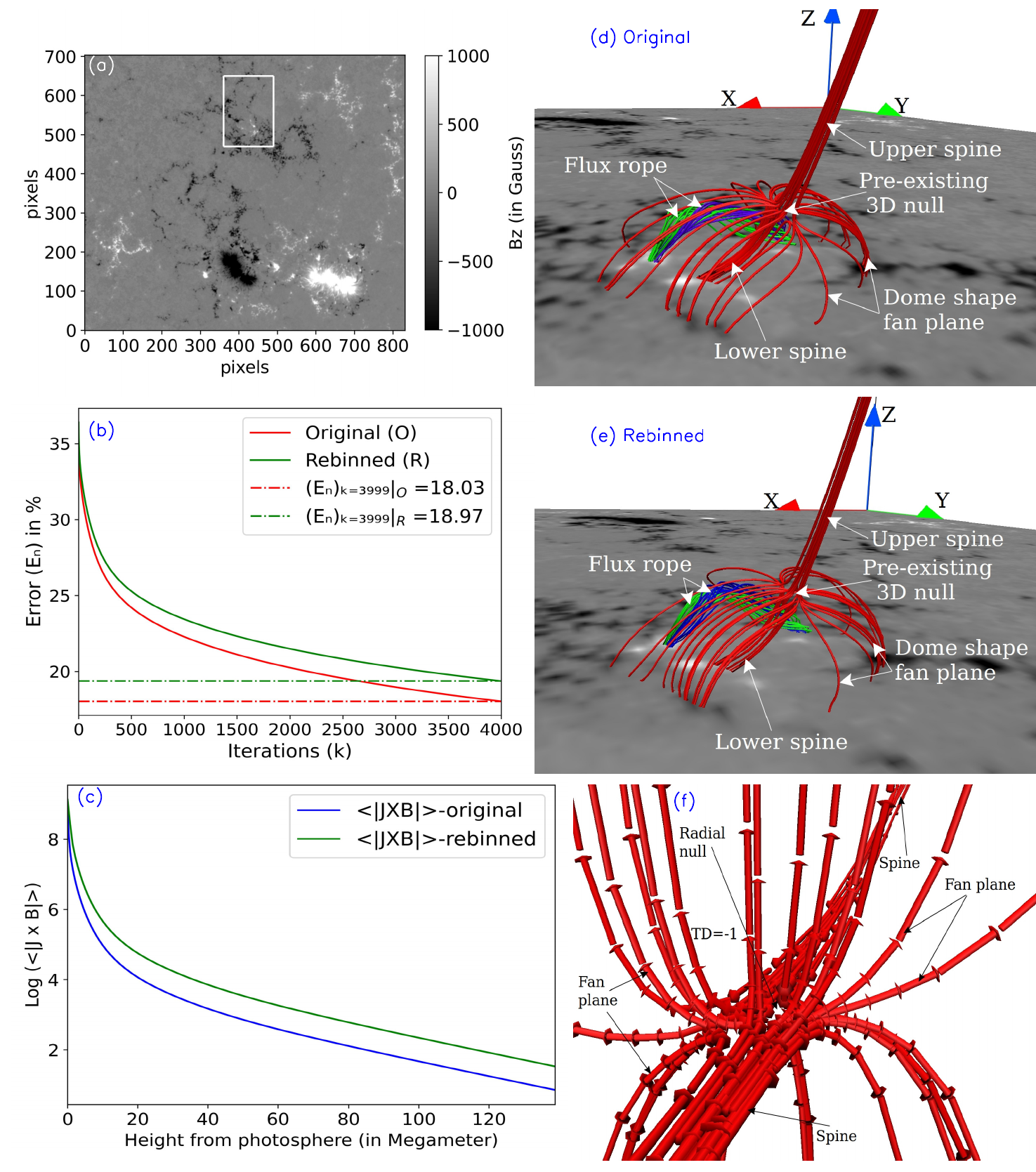}
\caption{Panel (a) depicts the magnetogram used for extrapolation with ROI marked by a white rectangular box. The variation in $E_{n}$ and average of $|J \times B|$ for both original (O) and Rebinned (R) resolution are shown in panels (b) and (e), respectively. Plots show no significant differences. Panels (d) and (e) depict the 3D magnetic null point and flux rope topology found co-located with flare and jet activity for original and rebinned extrapolation, respectively. Both show similar magnetic topology. In panel (f), the fan field lines (in red) of 3D null are directed away from null, making topological degree $-1$ and plotted from rebinned extrapolation.}
\label{fig:extrapolation_res}
\end{figure}

The extrapolated coronal magnetic field $\mathbf{B}$ corresponding to the photospheric magnetogram shown in panel (a) of Fig. \ref{fig:extrapolation_res} is obtained by employing the non-force-free field (NFFF) extrapolation technique developed by \citet{2008SoPh..247...87H} and \citep{2008ApJ...679..848H, 2010JASTP..72..219H}. The magnetogram in panel (a) is obtained at $03:00$ UT on November $10, 2022$ by cropping the observed HMI magnetogram of size ($1374 \times 716$) pixels to a size of ($832 \times 704$) pixels spans over ($301.6 \times 255.2 $) Mm in the x, and y axes, respectively. The net flux balance is maintained during cropping. The white (black) patches represent the positive (negative) polarities, respectively and the strength of $Bz$ is shown by the color bar. The ROI (flare and jet locations) is marked by the white rectangle. The HMI magnetogram used in this study is obtained from the `hmi.sharpcea720s' data series, which gives the vector magnetic field on a Cartesian grid that has been remapped using a Lambert cylindrical equal-area (CEA) projection, after which it is transformed in heliographic coordinates \citep{2014SoPh..289.3549B}.

The utilized NFFF extrapolation model is based on the minimization of the total energy dissipation rate, described in \citet{2007SoPh..240...63B}. The NFFF $\mathbf{B}$ obeys a double-curl Beltrami equation for which a solution can be given as follows,
\begin{equation}
\textbf{B} = \textbf{B}_{1} + \textbf{B}_{2} + \textbf{B}_{3};\qquad \nabla \times \textbf{B}_{i} = \alpha_{i} \textbf{B}_{i},
\end{equation}
where $i = 1, 2, 3$ \citep{2008SoPh..247...87H, 2008ApJ...679..848H}. Here, each sub-field $\textbf{B}_{i}$ represents a linear force-free field (LFFF) characterized by specific constants $\alpha_{i}$. Without loss of generality, a selection $\alpha_{1} \neq \alpha_{3}$ and $\alpha_{2} = 0$ can be made, rendering $\textbf{B}_{2}$ a potential field. Subsequently, an iterative approach is employed to determine the optimal pair $\alpha = {\alpha_{1}, \alpha_{3}}$, which finds the pair by minimizing the average deviation between the observed ($\textbf{B}_{t}$) and the calculated ($\textbf{b}_{t}$) transverse field on the photospheric boundary. Effectively, the metric $\textnormal{E}_{n}$ is given by
\begin{equation}
\textnormal{E}_{n} = \bigg(\sum_{i=1}^{M}|\textbf{B}_{t,i} - \textbf{b}_{t,i}|\times |\textbf{B}_{t,i}| \bigg)\bigg / \bigg(\sum_{i=1}^{M}|\textbf{B}_{t,i}|^{2} \bigg),
\end{equation}
where $M = N^{2}$ represents the total number of grid points on the transverse plane is iteratively minimized \citep{2018ApJ...860...96P}. To achieve an optimal value of $E_{n}$, a corrector potential field to $\textbf{B}_{2}$ is further derived from the difference transverse field, i.e., $\textbf{B}_{t} - \textbf{b}_{t}$, and added to the previous $\textbf{B}_{2}$ in anticipation of an improved match between the transverse fields, as measured by the metric $E_{n}$. The grid points are weighted with respect to the strength of the observed transverse field to minimize the contribution from weaker fields (see \citet{2008SoPh..247...87H}; \citet{2010JASTP..72..219H}, for further details). 

Two sets of NFFF extrapolations are performed: one using the original HMI resolution of $0.5^{"}pixel^{-1}$ over ($832 \times 704 \times 384$) pixels spanning a physical volume of ($301.6 \times 255.2 \times 139.2 $) Mm, and another using a re-binned version of the same physical volume but reduced to ($416 \times 352 \times 192$) pixels (with resolution $1^{"}pixel^{-1}$) in x, y, and z axes, respectively. The following analyses confirm the extrapolated magnetic field of the original and reduced resolution are almost identical. The percentage change in metric $E_{n}$ for original (O) (red solid curve) and rebinned (R) (green solid curve) resolution shows an initial sharp decrease but eventually gets saturated to values $18.03$ for O (red horizontal dash line) and $18.97$ for R (green horizontal dash line) with $4000$ iterations as shown in panel (b) of Figure \ref{fig:extrapolation_res}. Consequently, the iteration has stopped at $k=4000$ to limit computational cost. The logarithmic variation of horizontally averaged Lorentz force density with height for both the cases (blue curve is for original and green curve is for rebinned resolution) is shown in panel (c) and suggests that the Lorentz force is non-zero near the photosphere but almost vanishes at coronal heights, as generally expected \citep{2020ApJ...899...34L, 2020ApJ...899L...4Y}. This trend is fully consistent with the findings of \citet{2018ApJ...869....2B}, who quantified the height dependence of the Lorentz force in a 3D MHD simulation of an active region and showed that the $\mathbf{J}\times\mathbf{B}$ term is significant in the lower atmosphere but decreases rapidly with height, becoming negligible in the corona.

A set of selected MFLs are overlaid with the extrapolated magnetograms in the panels (d) and (e), corresponding to the extrapolation with the original resolution and the rebinned one, respectively, at the ROI. In the extrapolated field, a 3D magnetic null and a flux rope (field lines drawn in green and blue) under the dome of the 3D null is found to be present in the ROI. Both the plots show similar magnetic flux rope and 3D null topology with well defined spine and fan planes, the null being located at a height of $\approx 7.25$ Mm---corroborating the rebinned extrapolation retains the same topology of the original extrapolation. The magnitude of the twist within the flux rope is found to be around $-1.2$ (see Sect. \ref{sec:MHD_simulations} for details.) The field lines are drawn in red at the null. The lower spine of 3D null is anchored to a positive (white) polarity patch whereas the upper spine is open (may be closed at a faraway positive polarity region) and fan plane field lines are anchored to surrounding negative (black) polarity patches of the magnetic field making a dome-shaped fan plane. Here, the red, green, and blue arrows (emanating from a common point) represent the x, y, and z axes, respectively and in all such figures appearing throughout the paper. The panel (f) zooms at the location of the null and shows fan plane field lines (in red) to be directed away from the null, making its topological degree $-1$.

The 3D magnetic nulls are detected using an updated version \citep{2024PhyS...99g5017M, 2024PhPl...31g4502M} of the original trilinear null detection technique developed by \cite{trilinear_chitti} with details in \citet{2007PhPl...14h2107H}. The updated version can additionally determine the topological degree and the type of null based on the sign and type of the eigenvalues of the Jacobian matrix at the null. The field lines drawn at the 3D null are emanating radially (panel f) and all eigenvalues of the Jacobian matrix ($\nabla \mathbf{B}|_{null}$) at the null are real, confirming this is a radial null point. This radial null is denoted as the ``Pre-existing 3D null" to distinguish it from spontaneously generated 3D nulls found during simulation.

The theory of 3D nulls is well-established and the following properties are worth mentioning. A straightforward rearrangement of the ideal induction equation for an incompressible fluid leads to the following equation,
\begin{equation}
\label{dbdt}
\frac{d\textbf{B}}{dt}=0,
\end{equation}
at the null point \citep{doi:10.1063/1.871778}; the $d/dt$ being the Lagrangian derivative. Consequently, a 3D null maintains identity during evolution, enabling its tracing in numerical simulations. Moreover, the net topological degree of a system consisting of a number, $\textnormal{N}_{0}$ (initially), of nulls is defined by \citet{1992JCoPh..98..194G} and \citet{2005LRSP....2....7L} as
\begin{equation}
\label{td}
\textnormal{D} = \sum_{\textnormal{N}_{0}}\textnormal{Sign (det} (\nabla \textnormal{B}|_{x_{\textnormal{N}_{0}}} )),  
\end{equation}
and remains conserved \citep{doi:10.1063/1.871778, 2022LRSP...19....1P} in any evolution. Any simulation which explores the generation or annihilation of nulls must satisfy this stringent conservation. Additionally, nulls that enter or exit the computational domain need to be accounted for as they may seemingly violate the conservation of topological degree.

\section{MHD model and numerical setup}\label{sec:numerical_model_setup}
The simulation is carried out using EULAG-MHD \citep{SMOLARKIEWICZ2013608} which approximates the magnetofluid to be thermodynamically inactive, incompressible and having perfect physical electrical conductivity. The dimensionless governing equations are
\begin{equation}
 \frac{\partial \textbf{v}}{\partial t} + (\textbf{v} \cdot \nabla) \textbf{v} 
= -\nabla p + (\nabla \times \textbf{B})\times \textbf{B} + \frac{1}{R^{A}_{F}} \nabla^{2} \textbf{v} ,
\label{equ:mom_bal_eq}
\end{equation}

\begin{equation}
\nabla \cdot \textbf{v} = 0,
\label{eq:incom_con_eq}
\end{equation}

\begin{equation}
\frac{\partial \textbf{B}}{\partial t} = \nabla \times (\textbf{v}\times \textbf{B}),
\label{eq:ind_eq}
\end{equation}

\begin{equation}
\nabla \cdot \textbf{B} = 0
\label{eq:sol_con_eq}
\end{equation}

\noindent achieved with 
\begin{eqnarray}
\label{normalization}
&& \textbf{B}\rightarrow \frac{\textbf{B}}{B_{0}}, \textbf{v}\rightarrow \frac{\textbf{v}}{V_{A}}, L \rightarrow \frac{L}{L_{0}}, t \rightarrow \frac{t}{\tau_{a}}, p \rightarrow \frac{p}{\rho_{0}V^{2}_{A}},
\end{eqnarray}
where $R^{A}_{F} = (V_{A} L)/\nu$ is an effective fluid Reynolds number with $V_{A}$ as the Alfv\'en speed and $\nu$ as the kinematic viscosity. The $B_{0}$ and $L_{0}$ are characteristic values of the system under consideration, whereas $\rho_{0}$ represents the constant mass density. Although not strictly applicable in the solar corona, the incompressibility is invoked in other works also \citep{1991ApJ...383..420D, 2005A&A...444..961A}. With details in \citet{SMOLARKIEWICZ2013608} (and references therein), salient features of the EULAG-MHD model are highlighted here. The model uses the spatiotemporally second-order accurate, non-oscillatory, forward-in-time, multidimensional, positive-definite advection transport algorithm (MPDATA) \citep{https://doi.org/10.1002/fld.1071}. The advantage of MPDATA in relation to this work is two-fold. First, the momentum transfer equation (Eqn. \ref{equ:mom_bal_eq}) and induction equation (Eqn. \ref{eq:ind_eq}) are both solved in the Newtonian form i.e. Lagrangian derivatives of dependent variables and the associated forcings constituting the left- and right-hand sides of the two equations; see Section 2.1 in \citet{SMOLARKIEWICZ2013608} for details. Identity of null preservation is then assured since the forcing terms of the induction equation (Eqn. (\ref{eq:sol_con_eq})) vanish at the nulls to the accuracy of the field solenoidality, which is highly preserved in the computation \citep{SMOLARKIEWICZ2013608}. 
Secondly, the proven dissipative nature of the MPDATA is intermittent and adaptive to the generation of under-resolved scales in field variables for a fixed grid resolution \citep{2003PhFl...15.3890D, 2011ApJ...735...46R, 2016AdSpR..58.1538S}. Consequently, the MPDATA generates a residual dissipation which acts only at the locations where fields are spatially under-resolved and sustains the monotonic nature of the solution in advective transport. Such a localized dissipation of magnetic field triggers magnetic reconnection. Notably then, the reconnection is only in the spirit of Implicit Large Eddy Simulations (ILESs) that mimics the action of explicit subgrid-scale turbulence models, whenever the concerned advective field is under-resolved, as described in \citet{2006JTurb...7...15M}. Such ILESs carried out with the model were successful in replicating regular solar cycles by \citet{2010ApJ...715L.133G} and \citet{2011ApJ...735...46R} along with the rotational torsional oscillations which were further characterized and analyzed in \citet{2013SoPh..282..335B}. The simulations carried out here also rely on this ILESs's property to initiate magnetic reconnections already shown by \citet{2016ApJ...830...80K}.

The above delegation of the entire magnetic diffusivity to ILES has its own advantages and limitations. Although the localized and intermittent residual dissipation maximizes the effective Reynolds number away from the reconnection sites while reducing the computational cost \citep{2008JFM...606..239W, 2009JCoPh.228...33S} , nevertheless, it does not directly associate the electric field with the current density through Ohm's law. Without this association, it is difficult to estimate the magnetic Reynolds number. Further, being spatio-temporally intermittent, residual dissipation can only be meaningfully quantified in the spectral space where, akin to the eddy viscosity of explicit subgrid-scale models for turbulent flows, it is effective only on the shortest modes admissible on the grid \citep{2003PhFl...15.3890D}, particularly, at the vicinity of steep gradients in simulated fields. Such a calculation is beyond the scope of this paper.

\subsection{Numerical Setup} \label{sec:numerical_setup}

The simulation is performed with the initial magnetic field obtained from NFFF extrapolation, and the initial state is motionless ($\textbf{v} = 0 $). The non-zero Lorentz force associated with the extrapolated field pushes the plasma to generate dynamics. The mass density is set to $\rho_{0}=1$. The simulation is carried out on a computational grid of $416 \times 352 \times 192$ pixels, corresponding to a spatial domain x $\in \{-0.590, 0.590\}$, y $\in \{-0.50, 0.50\}$, and z $\in \{-0.272, 0.272\}$ in a Cartesian coordinate system, spanning a physical volume of approximately $301.6 \textnormal{Mm} \times 255.2 \textnormal{Mm} \times 139.2 \textnormal{Mm} $ in X-, Y- and Z- directions, respectively. The dimensionless spatial step sizes are $\Delta x$ = $\Delta y$ = $\Delta z$ $\approx$ 0.00284 ($\approx 725$ km) and the dimensionless time step is $\Delta t$ = $2 \times {10}^{-3}$ ($\approx 1.2$ s). The effective fluid Reynolds number ($R_{F}^{A}|_{\text{sim.}}$) is set to $8000$ (with $\frac{\tau_{a}}{\tau_{\nu}} = 1.25 \times 10^{-4}$), which is equal to $0.32 \times R_{F}^{A}|_{\text{cor.}}$. Where $R_{F}^{A}|_{\text{cor.}}$ is effective fluid Reynolds number for corona and its value is $25000$, calculated using typical value in the Solar corona, $L_{\text{cor.}} =100 \text{Mm}$, $V^{A}|_{\text{cor.}} = 1 \text{Mm s$^{-1}$}$, and kinematic viscosity $\nu|_{\text{cor.}} = 4 \times {10}^{9}$ \text{m$^{2}$ s$^{-1}$} (see page 791 of \citet{2005psci.book.....A}). Without any loss in generality, the reduced \text{R$_{F}^{A}$} can be realized as a reduction in computed Alfv\'en speed, \text{V$_{A}|_{\text{sim.}}$} $\approx$ 0.23 $\times$ \text{V$_{A}|_{\text{cor.}}$}. The Alfv\'en speeds are estimated with characteristic scales L$_\textnormal{sim.} = 139.2 \textnormal{Mm}$ for the simulation domain and L$_\textnormal{cor} = 100\textnormal{Mm}$ for typical corona. The results presented herein pertain to a run for $950 \Delta t$, which corresponds to an observation time of $\approx$ $19$ minutes. The reduced \text{R$_{F}^{A}$} slows down the dynamics and does not affect the reconnection mechanism or its consequence, reducing the computational costs and making data-based simulations computationally less costly---realized by \citet{2016NatCo...711522J}. Nevertheless, such a reduction in Alfv\'en speed will directly affect the wave dynamics, which is overlooked in this paper in favor of the reconnection dynamics.

\section{Simulation results}\label{sec:MHD_simulations}

\subsection{Pre-jet reconnection}
\begin{figure}
\includegraphics[width=0.97\textwidth]{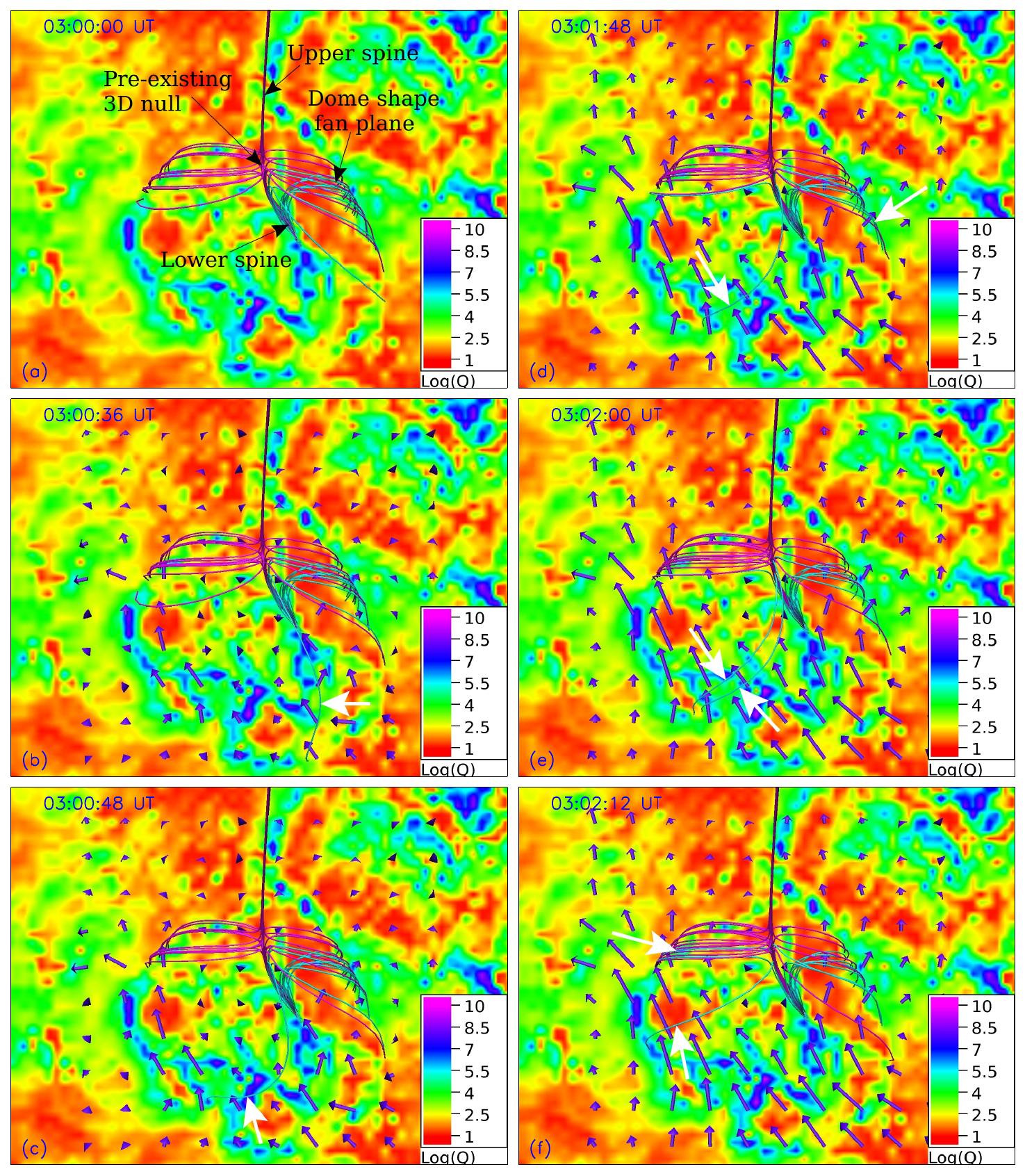}
\caption{This figure illustrates the slipping of the fan field lines (in cyan) through the plasma-i.e., slip-reconnection. The cyan field lines are traced from a fixed initial location for the integration. Plasma flows are drawn by blue arrows at the bottom boundary ($z = 0$) in all panels except (a), where no flow is present. In each panel, $\log(Q)$ is overlaid with the field lines. The fan field lines marked with white arrows move in a direction different from the plasma flow and traverse through high Q-value regions. The corresponding animation is available in the final HTML version with $t\in\{03:00, 03:03\}$ UT where the slippage of the field lines can be better visualized. The real time of the animation is $08$ s.}
\label{fig:qsl_reco}
\end{figure}

The data-constrained MHD simulation is carried out with the extrapolated magnetic field as the input while the initial velocity is kept zero. Moreover, $B_z$ is kept constant at the bottom boundary since the photospheric magnetic flux remains almost constant. The dynamics is generated as the Lorentz force acts on the plasma. The evolution depicted in Fig. \ref{fig:qsl_reco} demonstrates the slippage of foot-points of the fan field lines of the pre-existing 3D null---slip reconnection.
\begin{figure}
\includegraphics[width=\textwidth]{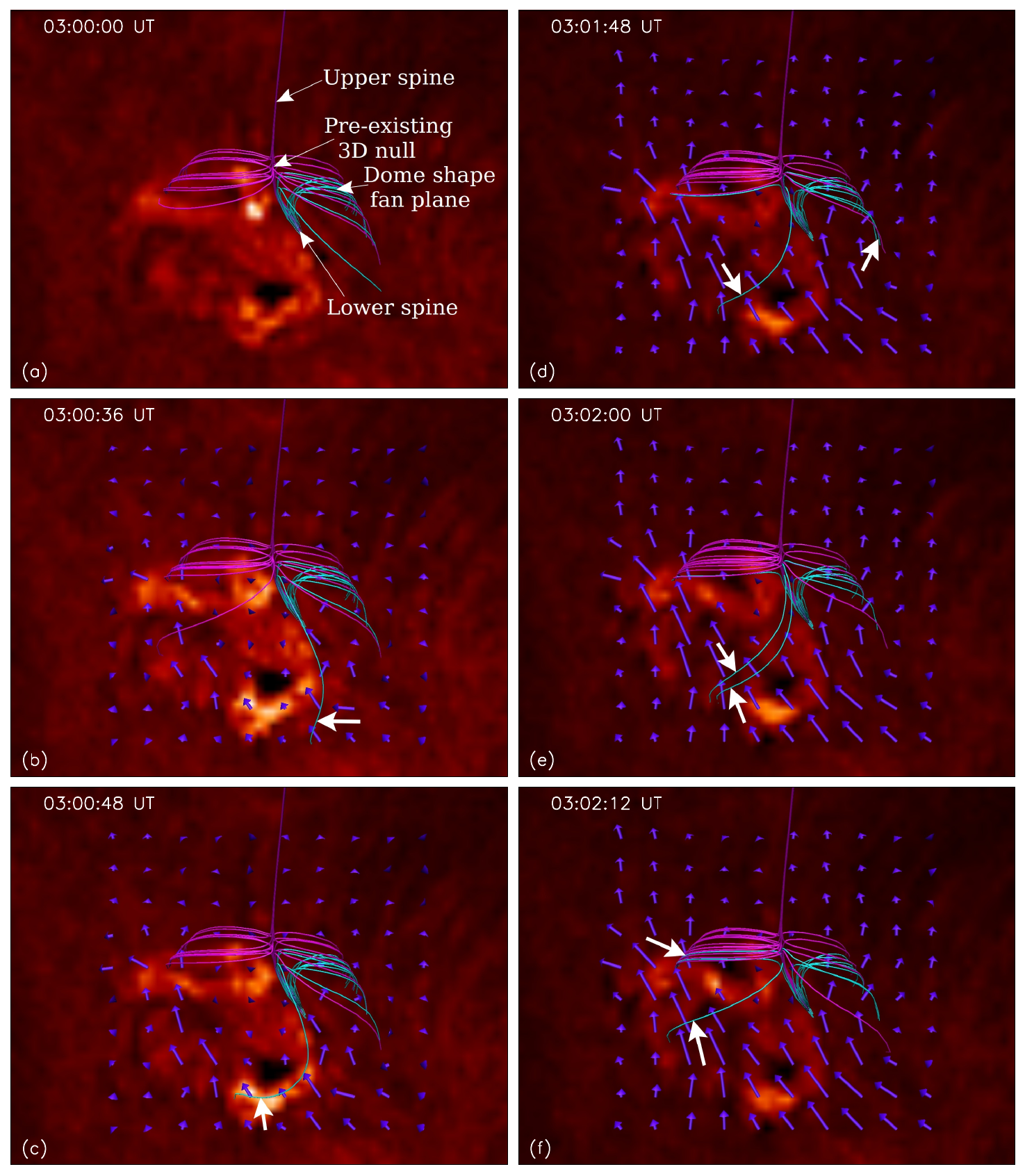}
\caption{The figure illustrates the brightening associated with slip-reconnection of the fan field lines of the pre-existing 3D null. Pink and cyan field lines are drawn and traced in same manner as described in Fig. \ref{fig:qsl_reco}. The AIA $304$ \AA~channel is overlaid to capture the signatures of slip-reconnection. With evolution, white arrows marked cyan fan field lines trace the bright patches (cf. panels (b)-(c) and panels (d)-(f)). An animation is available covering time $\in \{03:00, 03:03\}$ UT for a better visualization of brightening associated with slip reconnection. The real time of the animation is $08$ s.}
\label{fig:slip_reco_with_304}
\end{figure}
\noindent The two sets of field lines (pink and cyan) are drawn and overlaid with $\log(Q)$, where Q is the squashing factor defined in \citet{1996A&A...308..643D, 1997A&A...325..305D}. The magnitude of $\log(Q)$ are shown by the colorbar, where red (pink) represents the lowest (highest) value. The pink field lines are drawn at pre-existing 3D null to clearly illustrate its topology and cyan field lines are drawn to demonstrate slip-reconnection. The starting points of these cyan field lines for integration are chosen near the footpoint of the spine field line at the bottom boundary ($z = 0$ plane), and these points are kept fixed with time. Therefore, one end of each cyan field line is always traced from the same location. The initial non-zero Lorentz force generates plasma flow shown by blue arrows (in $z = 0$ plane) in all panels except panel (a) - which represents the plasma flow. With the evolution, a cyan field line marked by a white arrow (panel (b)), connects to a new location with a higher Q value (panels (c)). This is because of the steep gradient present in the magnetic field near the null point and along the quasi-separatrix layers (QSLs), therefore, parts of the cyan field lines diffuse and slip through the plasma parcels, causing their connectivity to change. The direction of plasma flow is different from that of the field lines, a breakdown of flux-freezing is suggested here and confirms the motion of the field lines are entirely due to non-ideal effects. Similarly, the other cyan field lines (marked by white arrows) also show the slip reconnection and passing through high Q regions (cf. panels (d)–(f)). To avoid clutter, the upper and lower spines, the dome-shaped fan-plane, and the pre-existing 3D null are marked in panel (a) only. An animation (movie-3) is provided for a better visualization of these slip reconnections which spans $\in\{03:00, 03:03\}$ UT.

As a consequence of the slip reconnection, when the plasma associated with these cyan fan field lines enters into the denser chromosphere, it produces chromospheric brightening \citep{2009ApJ...700..559M, 2013ApJ...774..154P}. The Fig. \ref{fig:slip_reco_with_304} demonstrates the brightening associated with slip reconnection of the fan field lines of the pre-existing 3D null overlaid with AIA $304$ \AA. The field lines are traced in the same manner as described for Fig. \ref{fig:qsl_reco}. A cyan fan field lines marked with white arrows (panel (b)) trace the bright patch as it evolves (panel (c)). Similarly, the other marked cyan field lines also trace the brightenings associated with slip reconnection (panels (d)-(f)). An animation (movie-4) is available spanning time $\in\{03:00, 03:03\}$ UT which provides a better visualization of field lines traversing through bright patches.
 
\begin{figure}
\includegraphics[width=\textwidth]{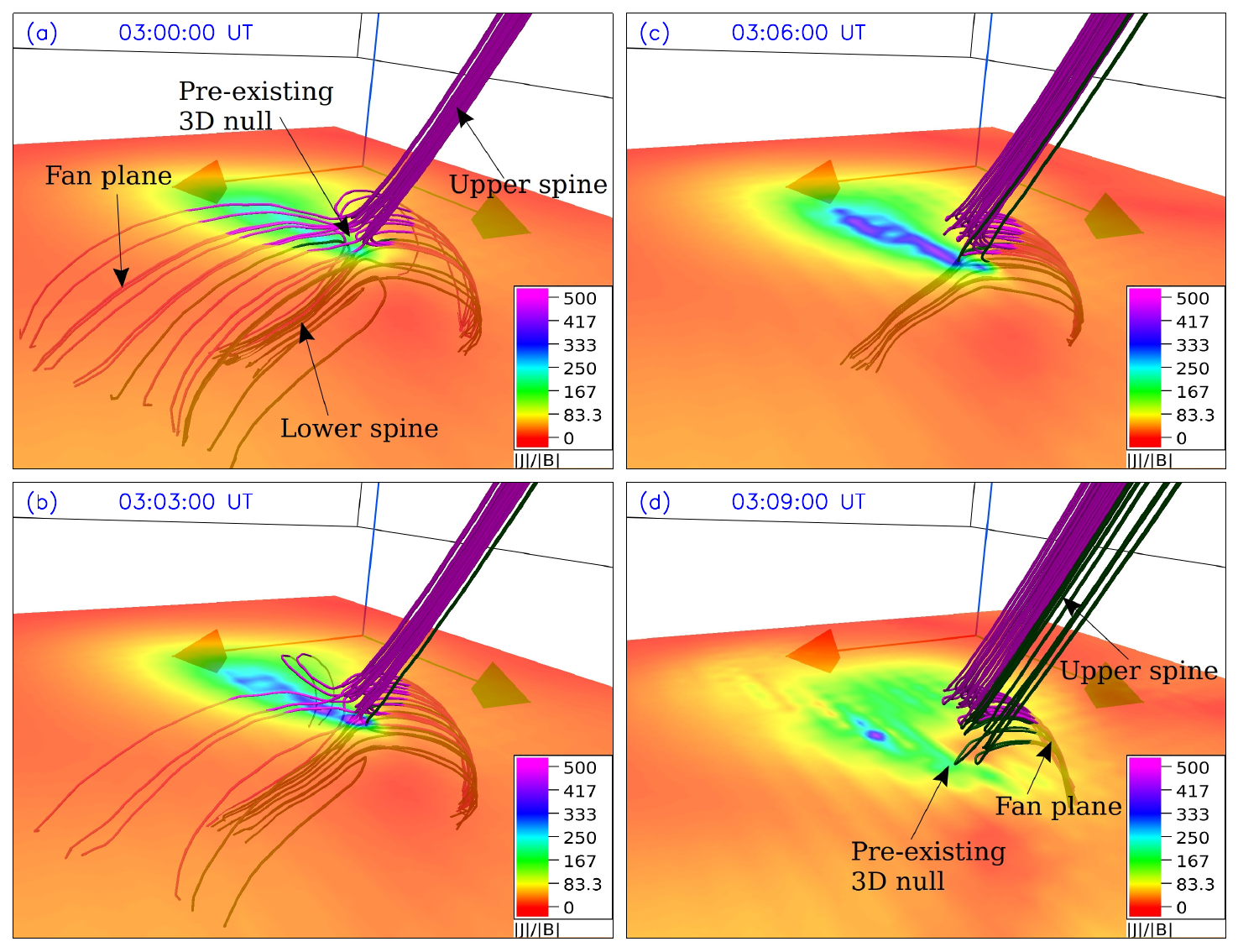}
\caption{The figure depicts null point reconnection overlaid with the probe of $|\mathbf{J}|/|\mathbf{B}|$ values, represented by the color bar shown in each panel. Initially, all green and two pink field lines are part of the lower spine whereas the all other pink field lines are part of the lower spine (panel (a)). Both sets (green and pink) of field lines are traced from a region away from the non-ideal, i.e., in the ideal region. With the evolution, all green and two pink field lines change their connectivity in the enhanced $|\mathbf{J}|/|\mathbf{B}|$ region and become part of the upper spine (cf. panels (a)-(d)). The variation in $|\mathbf{J}|/|\mathbf{B}|$ is seen accordingly. An accompanying animation covering time $\in \{03:00, 03:09\}$ UT is available which provides a better visualization of null point reconnection. The real time of the animation is $11$ s.}
\label{fig:central_null_reco}
\end{figure}

Figure \ref{fig:central_null_reco} shows that the reconnection is associated with a current sheet that forms around the pre-existing 3D null point. The two sets of field lines (one set in pink and the other in green), are drawn in the ideal region (away from null) to demonstrate the reconnection. All the green and two pink field lines are nearby the lower spine and fan plane whereas all other pink field lines are near by the upper spine and fan plane (panel (a)). A slice of $\mathbf{|J|/|B|}$ is overlaid to identify the large value of $\mathbf{|J|/|B|}$ i.e., high gradients in the magnetic field. Initially at $03:00:00$ UT, a lower value is present at the null point (panel (a)). With the evolution, $\mathbf{|J|/|B|}$ increases and null point reconnection onsets, for example, in panel (b), one green field line changes its connectivity and becomes part of the upper spine---null point reconnection. Similarly, the two pink field lines which are part of lower spine initially change their connectivity through null point reconnection, and become part of the upper spine (c.f. panels (c)-(e)). With the subsequent evolution, all remaining green field lines also change their connection via null point reconnection, and become the upper part of the spine (cf. panels (b)-(d)). The change in $\mathbf{|J|/|B|}$ is seen accordingly. An animation (movie-5) is available which covers time $\in \{03:00:00, 03:09:00\}$ UT showing a better evolution of null point reconnection. The null point reconnection continues along with the flux rope/filament approaching the null point, as described in the following. In summary, Figs. \ref{fig:qsl_reco}, \ref{fig:slip_reco_with_304} and \ref{fig:central_null_reco} demonstrate that prior to the jet in our simulation and likely also in the observation, there is sustained reconnection at the null point.

\subsection{Blowout jet}

\begin{figure}[ht]
\includegraphics[width=\textwidth]{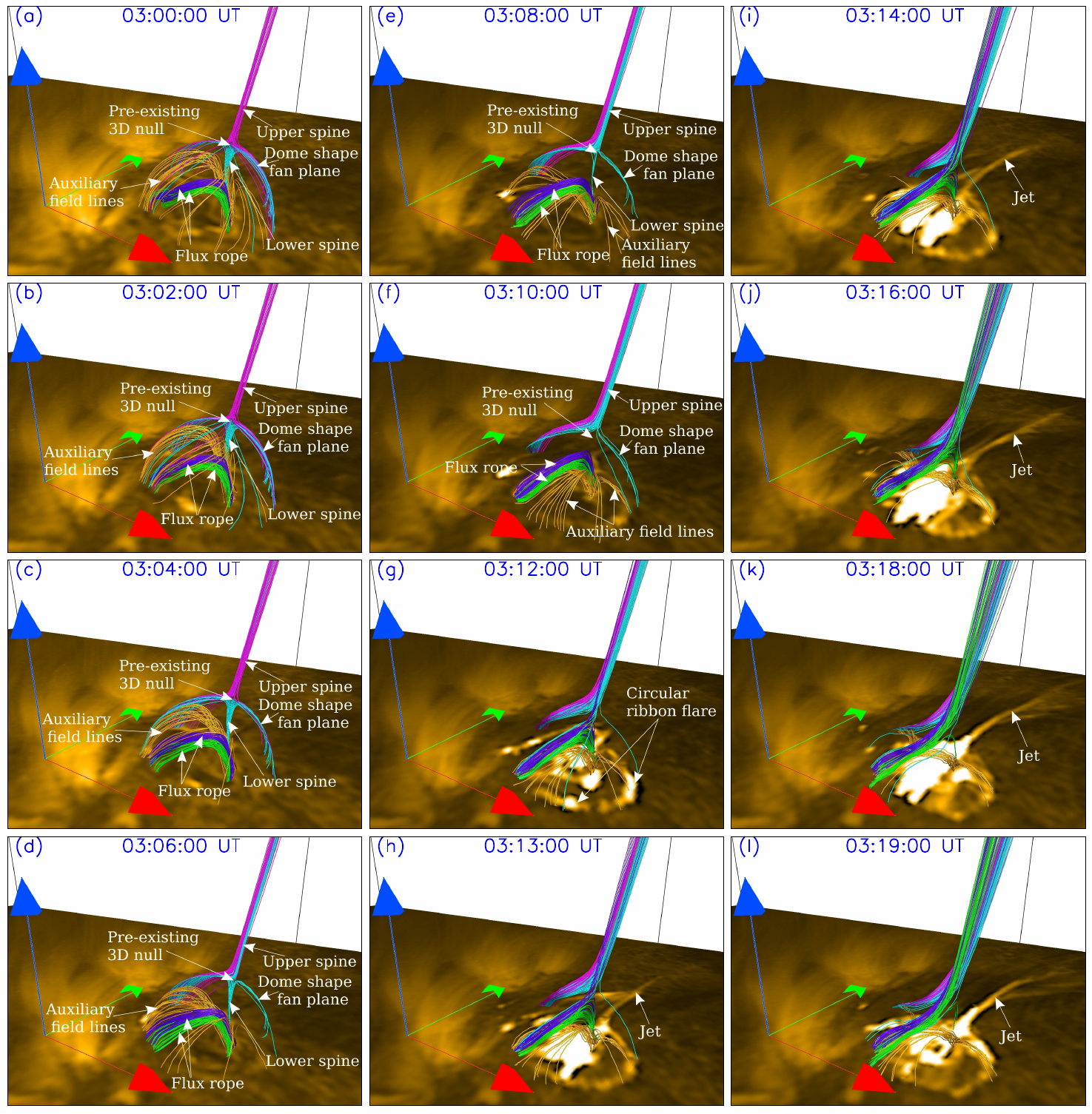}
\caption{The figure shows the evolution of field lines which consists of a flux rope, a 3D null point along with AIA observations, which manifested into a C $1.1$ class flare and launched a blowout jet. The flux rope field lines (in green and blue), auxiliary field lines (in yellow), lower (upper) spine, dome field lines in cyan (pink) are overlaid with AIA $171$ \AA~ channel. Initially, field lines adjacent to the lower spine becomes part of the upper spine of 3D null through the null point reconnection and the flux rope rises and approaches the null point (cf. panels (a)-(f)). With subsequent evolution, one leg of flux rope starts erupting via null point reconnection, resulting in a circular flare (panel (g)) and launching a jet (panel (h)). The eruption of flux rope continues and drives the jet (cf. panels (h)-(l)). An animation covering $\in \{03:00, 03:19\}$ UT is available for a better visualization of field lines, flare and jet. The real time of animation is $19$ s.}
\label{fig:eruption_of_flux_rope_aia_171}
\end{figure}

Figure \ref{fig:eruption_of_flux_rope_aia_171} follows the flux rope dynamics to explore the flare peak and the onset mechanism of the blowout jet in detail. The field lines show the skeleton of the 3D null (pink and cyan field lines), a set of auxiliary field lines (in yellow), and the flux rope (blue and green field lines). The bottom boundary represents images from $171$ \AA~channel of AIA. As the simulation evolves, the cyan fan field lines mimic the slip-reconnection, and reconnection at the 3D null point (shown in Figs. \ref{fig:qsl_reco} and \ref{fig:central_null_reco} explicitly). The brightening associated with null point reconnection is also seen but not shown here. Consequently, field lines adjacent to the lower spine reconnect through the fan surface via null point reconnection and align adjacent to the upper spine (cf. panels (a)-(f)). At the same time, the flux rope rises and approaches the null point (panel (f)). Initially, both ends of the flux rope are anchored to the bottom boundary. With subsequent evolution, blue field lines of the flux rope start reconnecting first and then green field lines start reconnecting through null point, making one end of the flux rope open. Consequently, the reconnection ramps up through null point and intensifies the brightening resulting in a C $1.1$ class flare around $03:12:00$ UT (panel (g)) and launches a blowout jet around $03:13:00$ (panel (h)). Furthermore, the eruption of remaining field lines has played a role in launching and driving the full blowout jet (cf. panels (h)-(l) of Figs. \ref{fig:eruption_of_flux_rope_aia_171} and animation (movie-6)). A similar agreement with brightening in other AIA channels is also found (not shown).

\begin{figure}
\includegraphics[width=0.95\textwidth]{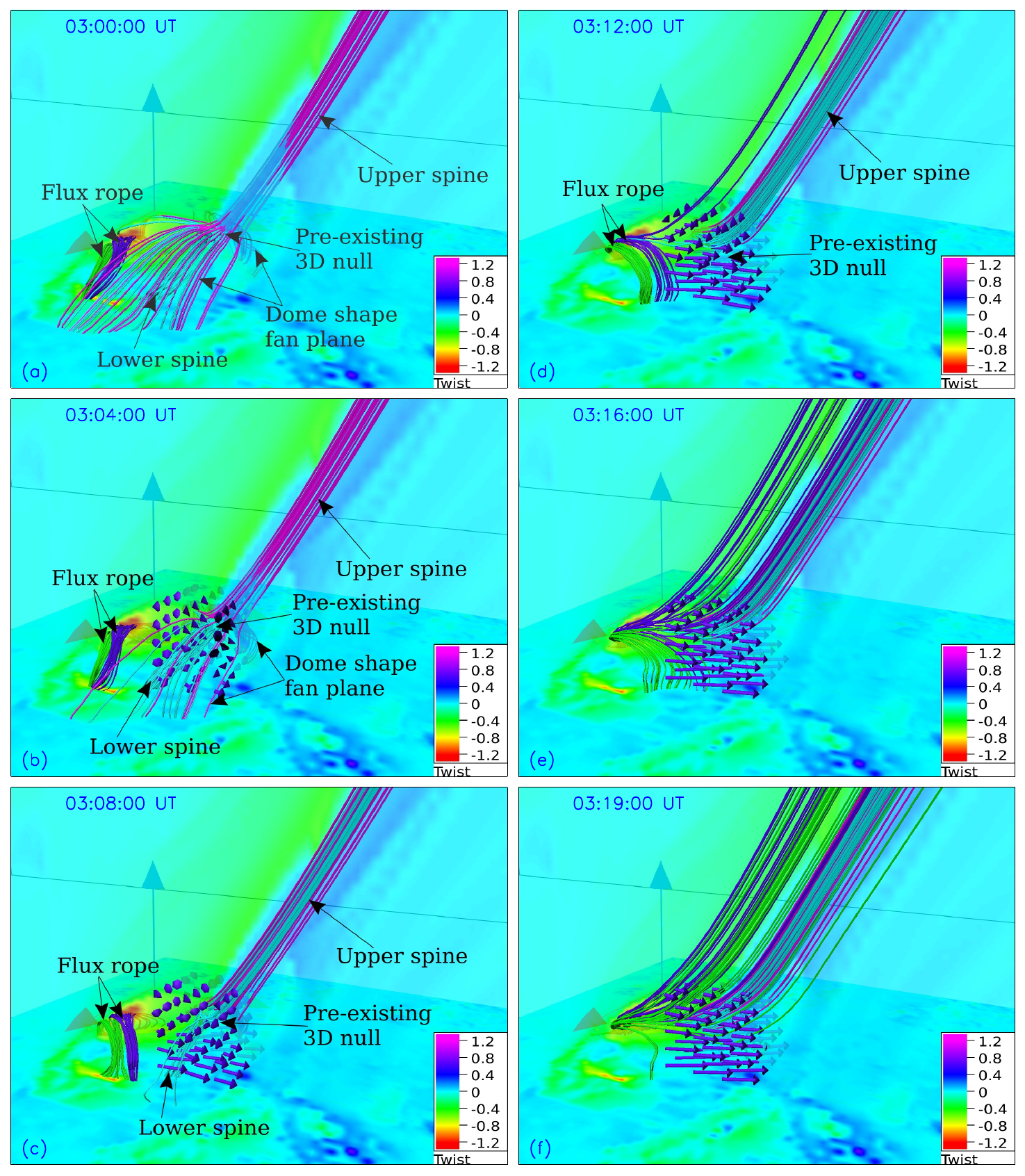}
\caption{The figure illustrates the evolution of flux rope overlaid with twist and flow resulting in a flare and launching a blowout jet. The lower (upper) spine, dome is shown through cyan (pink) field lines (panel (a)). The plasma flow is shown by blue arrows (panel (b) onward). Two slices (one horizontal and one vertical) of twist is overlaid with field lines. The color bar represents the magnitude of twist where red (pink) represents the negative (positive) twist, respectively. With the evolution, flux rope approaches the null point (cf. panels (a)-(c)) and one leg of flux rope erupts through the null point reconnection, the decrease in twist and plasma flow alignment is seen accordingly (panels (d)-(f)). An animation spanning time $\in \{03:00, 03:19\}$ UT is available for better visualization of evolution of twist, flux rope and flow. The real time of animation is $19$ s.}
\label{fig:eruption_of_flux_rope_twist}
\end{figure}

Inherent to a magnetic flux rope is its twist, and its interrelation during evolution merits further attention. Another important aspect of evolution is plasma flow. Fig. \ref{fig:eruption_of_flux_rope_twist} explores both. The green and blue field lines represent the flux rope (panel (a)) and their seed points are located in a region away from the pre-existing 3D null (from the non-ideal region) and traced in time from a fixed location with the evolution. Initially, around $03:00:00$ UT, the cyan field lines are adjacent to the lower spine and outline the inner dome of the null. While the pink field lines lie adjacent to the upper spine and outline the outer dome-shaped fan plane. In Fig. \ref{fig:eruption_of_flux_rope_twist}, the overlaid slices represent the twist which is calculated by the method described in \citet{2016ApJ...818..148L}. The color bar shown in the bottom right corner of the panels represents the twist parameter, where red (pink) represents the negative (positive) high twist. Twist around $-1.2$ is found around the flux rope (panel (a)). As can clearly be seen from the vertical slice, it remains almost constant with time until $03:12$ UT (panel (d)), from where it starts decreasing. This is expected as the rope begins open up through null point reconnection. This decrease continues with further reconnection, till one leg of the rope gets detached. With the evolution, the flux rope (the blue field lines first) started reconnecting at the pre-existing 3D null point, opening one end (panel (d)). Subsequently, the green flux rope field lines and the remaining blue field lines also open via reconnection at the null point (panels (d)-(f)). The plasma flow (shown by blue arrows) aligns accordingly (cf. panels (b)-(f)) and its magnitude (represented by length of the arrows) also increases with time.

Importantly, the rise, approach towards and reconnection of the flux rope with the open field coincides with a ramp up in null reconnection indicative of the observed C $1.1$ class flare, while the opening up of the flux rope field lines and transfer of twist to the open field coincides with jet-like plasma flows along the spire at the same time as the blowout jet is launched. This scenario of the simultaneous reconnection of the flux rope and the appearance of the blowout jet agrees with the findings of previous models for blowout jets \citep[e.g.][]{wyper2018,2013ApJ...769L..21A}.

\begin{figure}[ht]
\includegraphics[width=\textwidth]{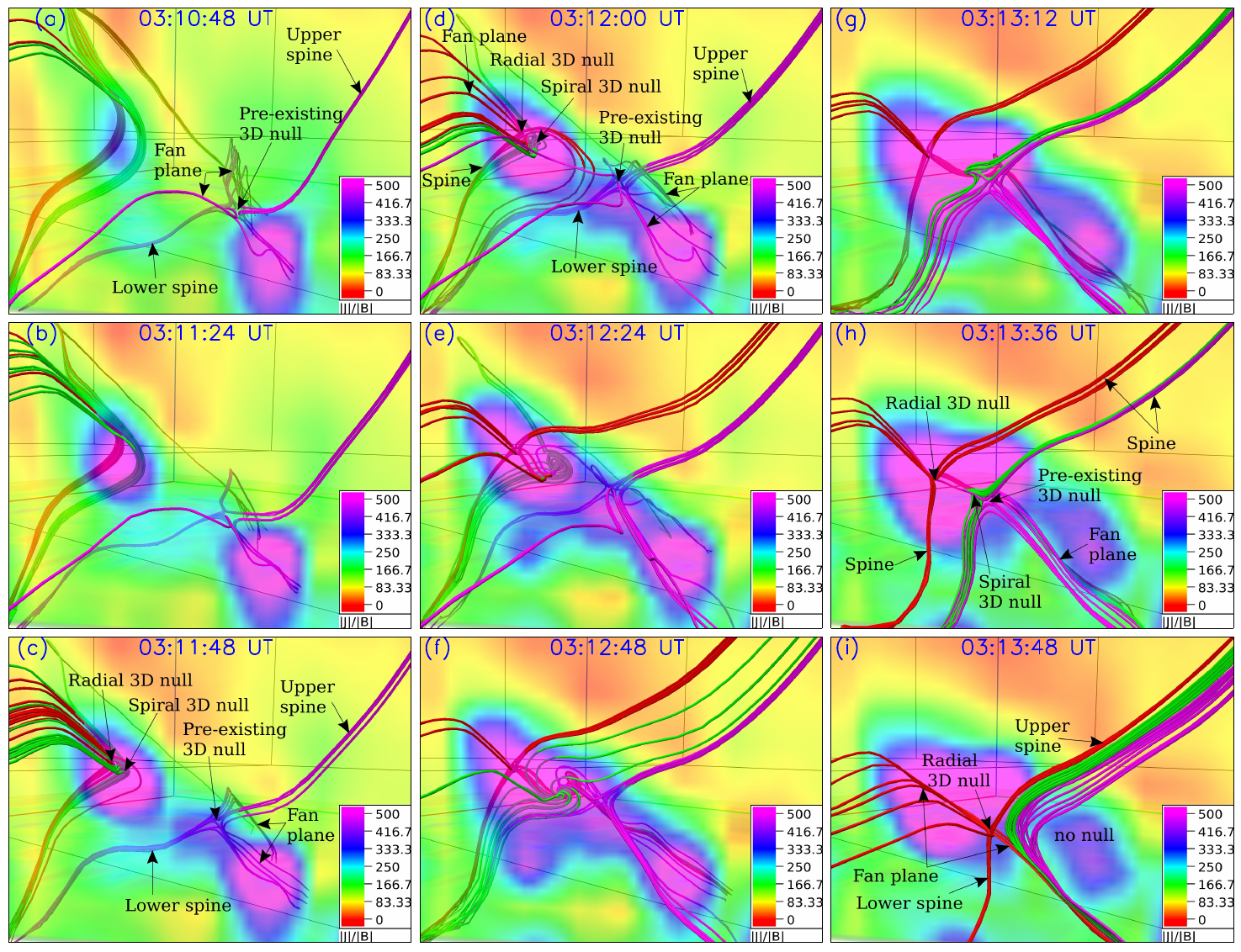}
\caption{The figure depicts the spontaneous generation of 3D nulls in a pair and subsequent annihilation in a pair. Initially, a pre-existing 3D null (pink field lines) is present (panel (a)). The red and green field lines are drawn to demonstrate the spontaneous generation of nulls. The quantity $\mathbf{|J|/|B|}$ is overlaid with field lines to identify its high value representing the non-ideal region. With the evolution, the green and red field lines develop an elbow shape, prominent in panel (b), $\mathbf{|J|/|B|}$ increases accordingly. Subsequent evolution shows the spontaneous creation of nulls in a pair which consists of a radial null and a spiral null (panel (c)). With further evolution, the spiral null recedes away from the radial null and simultaneously approaches the pre-existing null, and continues until annihilated in pair (cf. panels (d)-(i)). An animation (movie-7) covering time $\in\{03:10:48, 03:13:48\}$ UT is available showing a better visualization of spontaneous generation and annihilation of null. The real time of animation is $08$ s.}
\label{fig:spontaneous_null_reco}
\end{figure}

\subsection{Multiple null points}

The simulation also shows the spontaneous generation of 3D nulls in a pair around $t =03:11:48$ UT co-located with the flare and jet activity region, and is shown in Figure \ref{fig:spontaneous_null_reco}. The three sets (red, green, and pink) of field lines are drawn to demonstrate the spontaneous generation of 3D nulls, their evolution, and annihilation. The quantity $\mathbf{|J|/|B|}$ is overlaid and its value is shown in a color bar. The pink field lines are drawn at the pre-existing 3D null (having topological degree $-1$) and traced in time. The spine, fan plane, and pre-existing 3D null are marked by arrows (panel (a)) at $t = 03:10:48$ UT. The red and green field lines are drawn to demonstrate the spontaneous generation of 3D nulls. Initially, the curved field lines (the red, and green) are present and a smaller value of $\mathbf{|J|/|B|}$ is found (panel (a)). With evolution, the red and green field lines develop a prominent elbow shape (panel (b)), and the $\mathbf{|J|/|B|}$ increases accordingly. Further evolution shows the spontaneous creation of 3D nulls in a pair (panel (c)) at around $03:11:48$ UT. The pair consists of a radial and spiral null marked by arrows in panels (c) and (d). The subsequent evolution shows the spontaneously generated nulls recede away from each other and the spiral null approaches the pre-existing 3D null (c.f. panel (c)-(e)) and continues until they annihilate in a pair (panel (h)), leaving a single radial null once more (panel (i)). 

\begin{figure}[ht]
\includegraphics[width=0.985\textwidth]{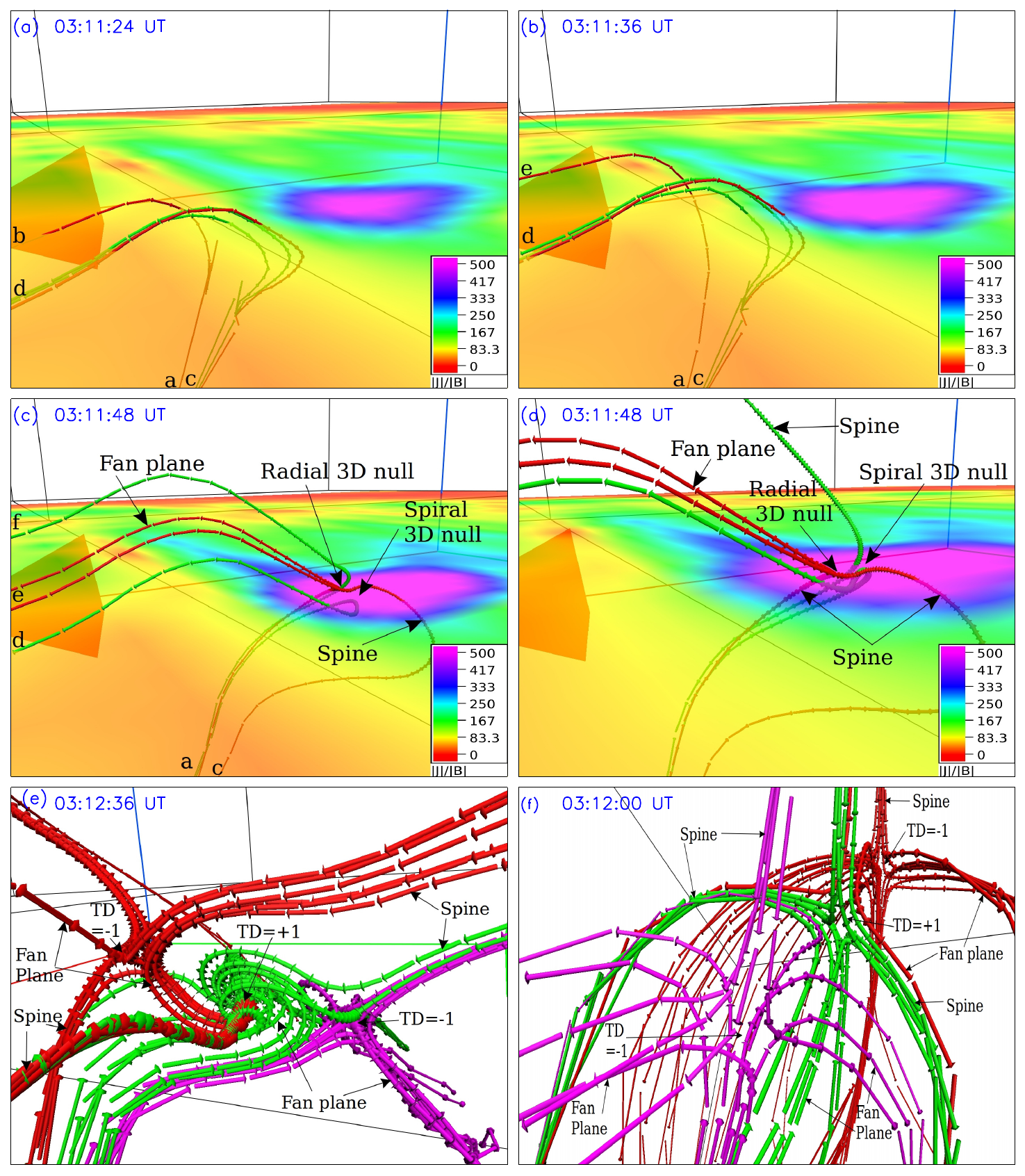}
\caption{Panels (a)-(c) represent the spontaneous generation of 3D nulls in a pair through magnetic reconnection. Two sets (red and green) of selected field lines are drawn away from the non-ideal region, advected with the plasma flow and traced in time. With the evolution, the red and green field lines change the connectivity---magnetic reconnection, generate the nulls in a pair (panel (a)) and the quantity $\mathbf{|J|/|B|}$ varies accordingly. Panel (d) shows a slightly zoomed and a different view angle of panel (c). The topological degrees of all 3D nulls are depicted in panels (e) and (f) obtained from simulation and extrapolation, respectively, and the overall net topological degree $-1$ is found to be conserved.}
\label{fig:reco_spontaneous_null_with_td}
\end{figure}

To explore the underlying field dynamics of spontaneous generation of the nulls, the evolution of two sets (red and green) of field lines are selectively drawn at the location away from the non-ideal region, i.e. in the region away from the high value of $\mathbf{|J|/|B|}$ (shown by a probe) is shown in panels (a)-(c) of Fig. \ref{fig:reco_spontaneous_null_with_td}. The field lines are advected with the plasma flow and traced in time. One red field line is connected to point `a' to `b' and second red field line along with two green field lines are connected to the point `c' to `d' (panel (a)). With the evolution, the red field line changes its connectivity from point `a' to `b' to the point `a' to `e' (panel b)). Subsequently, one green field line changes its connection from the point `d' to `c' to the point `d' to `a', the second green field line gets connected to the point `a' to `f'. Similarly, one red field line changes its connectivity from the point `c' to 'd' to the point `c' to `e' (c.f. panels (b)-(c))---telltale sign of magnetic reconnection. A pair of 3D nulls gets created with this change in connectivity of field lines shown and marked in panel (c). Panel (d) shows a slightly zoomed-in and a different view angle of the nulls. The topological details of the spontaneously generated along with pre-existing 3D nulls from simulation at around $03:12:36$ UT are shown in panels (e), where nulls, their spines, fan planes, and topological degrees are marked by arrows. The field lines in pink, green, and red are drawn at the pre-existing, spiral, and radial 3D null, respectively. The spine field lines (in pink) of pre-existing null are directed toward null, making the topological degree $-1$, the fan field lines (in green) are directed toward null, resulting in topological degree $+1$, and the spine field lines (in red) are directed toward null, making the topological degree $-1$. The spontaneous generation of nulls (shown by green and red field lines) in a pair having a complementary topological degree ($+1$ and $-1$) leads to an unaltered overall net topological degree $-1$.

To further establish the spontaneous generation of 3D null pair, the coronal magnetic field is extrapolated from the next available photospheric vector magnetogram at $03:12:00$ UT. The idea is to check if this null pair, which was absent in the previously constructed magnetic field at $03:00:00$, is present in this extrapolated field or not. Interestingly, two radial nulls, co-spatial with the spontaneously generated nulls are found in the extrapolated field. For clarity, panels (e) and (f) plot spontaneous nulls found in the simulation and the ones seen in the present extrapolation. Although collocated and having identical topological degree, the two sets differ in only their types; the simulated pair consists of a radial-spiral combination whereas both nulls in the extrapolated field are radial. The overall net topological degree is found to be $-1$ in both approaches, i.e., in the data-constrained MHD evolution and the NFFF extrapolation. Consequently, the net topological degree from both the evolution remains preserved.

\begin{figure}[ht]
\includegraphics[width=\textwidth]{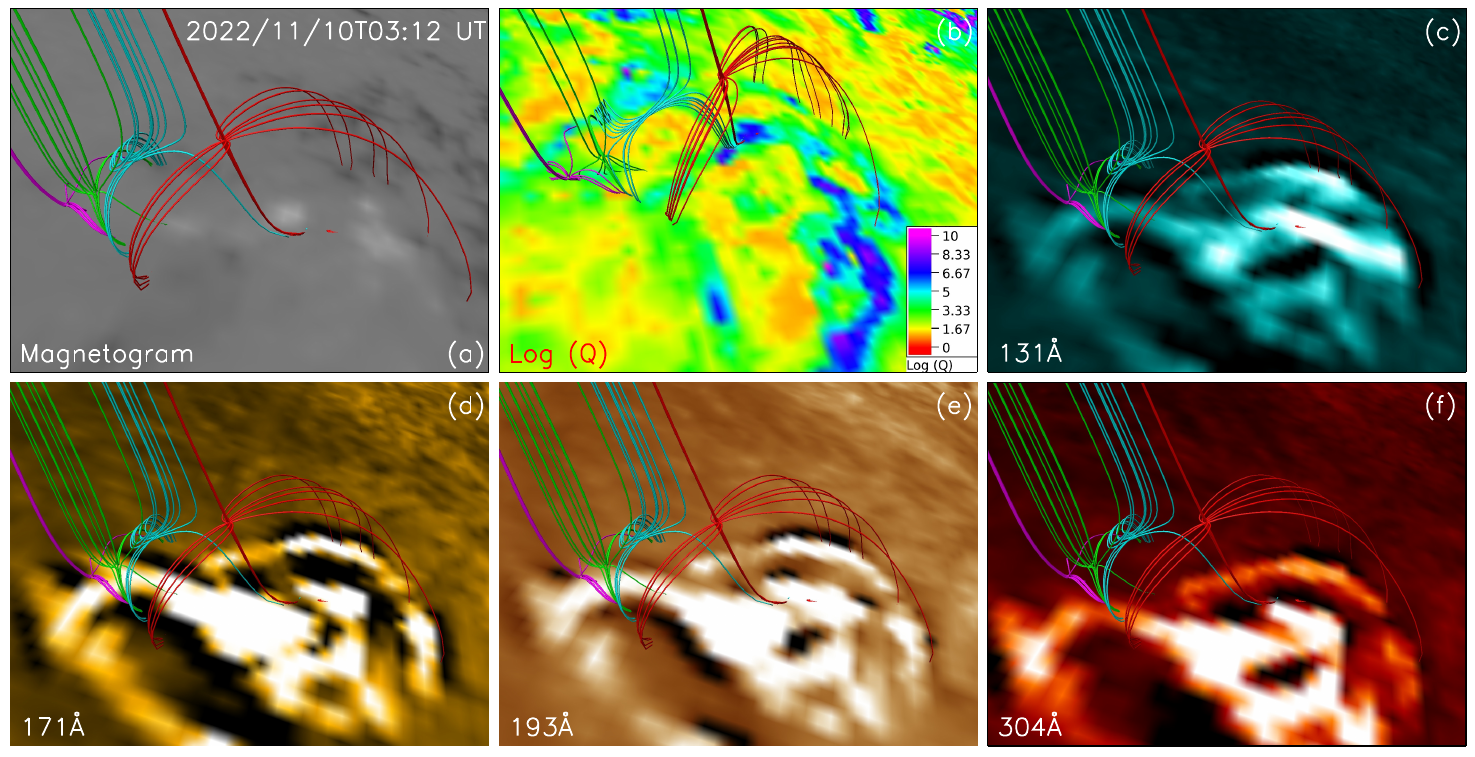}
\caption{The figure depicts the magnetic topology at $03:12$ UT found at the location of jet overlaid with $B_{z}$ component of magnetogram (panel (a)), $\log(Q)$ (panel (b)), AIA $131$ \AA~ (panel (c)), $171$\AA~ (panel (d)), and $304$ \AA~ (panel (f)). Overall, the magnetic topology found from extrapolation matches well with the different AIA observations.}
\label{fig:mag_312}
\end{figure}

To co-locate the nulls found in the extrapolated field, Fig. \ref{fig:mag_312} depicts the location of the nulls found in extrapolation with respect to the flare and jet region. The red field lines are drawn at the pre-existing null and green and pink field lines are drawn at the location of the additional two nulls. The field lines are overlaid with the $B_{z}$ component of the magnetogram in panel (a), $\log(Q)$ in panel (b), and different AIA channels. The overlaid magnetogram and AIA channels suggest that these nulls are co-located in the flare and jet region, like the ones found in the simulation. Consequently, this makes the findings more credible and realistic. The key difference between the two approaches is the distribution of current throughout the volume, with the simulation involving a concentration of current at the open-closed boundary, and hence the presence of a spiral null, compared to the extrapolation where the current is more globally distributed. Indeed, with higher resolution we might expect even more pairs of additional nulls while still maintaining the net topological degree \citep[e.g.][]{2016ApJ...827....4W}.

\section{Summary and discussions} \label{sec:summary}

A data-constrained Implicit Large Eddy magnetohydrodynamic (MHD) simulation has been performed to investigate the magnetic topology and onset mechanism of a C-class solar flare and blowout jet event observed by the Spatial Possibilistic Clustering Algorithm (SPoCA) suite with SPoCA number 29093 on $10$ November $2022$. The peak of the flare and onset of the jet are nearly simultaneous around $03:13$ UT. The coronal magnetic field at $03:00$ UT is obtained using a NFFF extrapolation model from the photospheric vector magnetogram to identify the magnetic topology. The extrapolation is performed for a cutout of ($832 \times 704$) pixels in x, and y directions with original HMI resolution and the height is $384$ pixels in z direction, which results in a volume of ($301.6 \times 255.2 \times 139.2$) Mm in x, y, and z axes, respectively. A 3D magnetic null (at a height of $\sim 7.25$ Mm above the photosphere) and flux rope/filament (under the fan dome of the null) is found to be co-located with the flare and jet region in the extrapolated field. To locate and determine the type and topological degree of the null, the null trilinear null detection tool has been employed. The located null is called a 'pre-existing 3D null' to distinguish it from spontaneously generated nulls. To save the computational cost, the original magnetogram is rebinned and extrapolated to a ($416 \times 352 \times 192$) pixels in x, y and z axes, respectively which corresponds to the same physical volume. The magnetic topology is found to be almost identical in both extrapolations. The magnetic field obtained from rebinned extrapolation is used as the initial magnetic field for the MHD simulation. The initial plasma flow is kept to zero and the non-zero Lorentz force drives the simulation.

The simulation shows the slippage of the fan plane field lines of the pre-existing 3D null through the plasma, traversing through high values of squashing factor (Q)---slip reconnection. The slip reconnection associated brightening is identified with AIA $304$ \AA observations and the fan field lines pass through bright patches in AIA observations. With evolution, the pre-existing null exhibits null point reconnection, identified with enhanced in $|\mathbf{J}|/|\mathbf{B}|$ near the pre-existing null, through which field lines near the lower spine are opened adjacent to the upper spine. With further evolution, the flux rope rises and approaches the null, and starts reconnecting through the null, which ramps up the reconnection, resulting in the C-class flare. The eruption and reconnection of the flux rope open up one of its legs and launch a blowout jet. Consequently, the plasma flows are collimated and aligned along the jet direction.

Further, it is found that during the evolution, the twist of the flux rope remains almost unchanged until it begins to reconnect at the null point. With onset of the reconnection, the twist within the section of the flux rope that has not yet reconnected starts to continuously decrease and achieves a value significantly smaller than its initial magnitude until it ultimately completely reconnects. Once connected to the open field, this twist propagates away along the open field helping drive the upward flows forming the blowout jet. Given its importance to the jet process, a relevant question is to understand the dependency of coronal field topology and twist on the initial extrapolated field. An important magneto-frictional simulation in this regard has been carried out by \citet{2024RAA....24b5007V}, quantifying differences in twist and topology of the evolved field depending on the initial magnetic fields obtained via potential and non-linear-force-free field (NLFFF) extrapolations. The results also indicate that field line twist is minimal with potential field extrapolation while maximal with the NLFFF. This is in accordance with the fact that the velocity of the induction equation being approximated by the Lorentz force in the magneto-frictional equations is non-linear. Non-linearity can generate twist during evolution of the magnetofluid which adds over the existing magnetic field. Similarly, the MHD equations are also non-linear and can generate and support additional twist. Contextually, it is important to note that the twist being measured by the field aligned current is maximum in NLFFF where current and magnetic field are fully aligned. In NFFF the same is not true, and hence, the magnitude of twist found here should be in between the values for the potential and NLFFF---although a direct comparison is beyond the scope of this paper. Regarding magnetic topology, the NLFFF and NFFF can agree well, for example, see \citet{2022SoPh..297...91A}. 

Notably, the simulation further revealed the spontaneous creation of a radial–spiral 3D null during the flare–jet evolution at around $03:11:48$ UT, with one null getting annihilated with the pre-existing null later. The NFFF extrapolation at $03:12$ UT also confirms the presence of two radial nulls in addition to the pre-existing 3D null. The additional nulls are found to be co-located with flare and jet region. The topological degrees of nulls of the extrapolated field are found to be identical to that of nulls found in the simulation, however, one null in the extrapolated field differed in its type from the spiral null found in the simulation because the latter concentrates current at the open-closed boundary, while the former distributes it more globally. The simulation, extrapolation and observational results are in close agreement, which lends strong credibility to the underlying magnetic reconnection process which launched the blowout jet and spontaneously generated the null, suggests the realistic topological changes in the coronal magnetic field.

The simulation provides results consistent with the theory, modeling, and observationally supported view of blowout jet initiation, linking magnetic reconnection processes to observed flare and jet dynamics. The identified slip-reconnection along high Q value field lines (i.e., QSLs) is consistent with previous reports of slipping motions of field lines in fan–spine topologies \citep{2006SoPh..238..347A, 2009ApJ...700..559M, 2013ApJ...774..154P}. The circular flare morphology observed here aligns with the morphology of 3D nulls and of classical null point reconnection models, where magnetic connectivity changes between magnetic domains separated by fan planes---null point magnetic reconnection, and consequently the energy release occurs \citep{1990ApJ...350..672L, 2009PhPl...16l2101P}. The eruption of one of the legs of the underlying flux rope via the null point reconnection leading to jet onset and alignment of the plasma flow along the spire is also consistent with the breakout model described by \citet{1999ApJ...510..485A, 2017Natur.544..452W, 2018ApJ...852...98W}, in which reconnection above sheared field lines or flux ropes removes overlying strapping field lines, allowing the sheared field lines/flux rope to rise and erupt. Further, the generation of nulls in pairs is in the agreement with the results shown in \citet{2016ApJ...827....4W} and the underlying cause of spontaneous null pair generation is consistent with the findings of the \citet{2023PhPl...30b2901M, 2024PhyS...99g5017M, 2024PhPl...31g4502M}, which requires local, non-ideal MHD effects \citep{2010ApJ...714..517E, 2011AdSpR..47.1508P,doi:10.1063/1.4896060} along with the conservation of the net topological degree as theorized by \citet{doi:10.1063/1.871778, 2022LRSP...19....1P}. Overall, this study highlights a multi-phase evolution sequence---slip-reconnection at QSLs, null point reconnection, rise, approach, eruption and null point reconnection of the flux rope producing a circular flare, and launching a blowout jet. A future work in this direction is to carry out a comparative study using compressible MHD simulations with explicit physical resistivity to quantify better the role of magnetic diffusivity in magnetic reconnection and the subsequent evolution of the coronal magnetic field. The present findings emphasize the importance of the magnetic topology and field line dynamics leading to a flare and an active region blowout jet, which can further help in understanding the energy release and mass transfer processes in the solar atmosphere.

\begin{acknowledgments}

The computations were performed on the Param Vikram-1000 High Performance Computing Cluster of the Physical Research Laboratory (PRL). This work is supported by the Department of Space, Government of India. We also wish to acknowledge the visualization software VAPOR (www.vapor.ucar.edu), for generating the relevant graphics. PW was also supported through a Leverhulme project grant.

\end{acknowledgments}

\bibliography{circ}{}
\bibliographystyle{aasjournalv7}

\end{document}